**Einstein and Gravitational Waves 1936-1938**

Galina Weinstein

15/2/16

Around 1936, Einstein wrote to his close friend Max Born telling him that, together with Nathan Rosen, he had arrived at the interesting result that gravitational waves did not exist, though they had been assumed a certainty to the first approximation. He finally had found a mistake in his 1936 paper with Rosen and believed that gravitational waves do exist. However, in 1938, Einstein again obtained the result that there could be no gravitational waves!

In 1916 and 1918 Einstein had done calculations of the gravitational waves in the weak field linearized approximation, where he made simplifying assumptions regarding the gravitational field. Einstein obtained his approximate solutions, of plane gravitational waves travelling at the speed of light, by introducing the harmonic coordinate condition into his field equations, a coordinate condition that led to linearized gravitational field equations. He found solutions of the weak field linearized field equations in a manner analogous to that of retarded potentials in electrodynamics.

Calculating the gravitational fields with the non-linearized gravitational field equations (i.e. finding an exact solution to the field equations of general relativity which describes plane gravitational waves) could yield different results. Indeed, it turned out that in the exact non-linear theory the solutions of plane gravitational waves did not satisfy the harmonic condition anymore and they contained singularities. Einstein therefore changed his mind with regard to gravitational waves and claimed they did not exist.

Around 1936, Einstein wrote to his close friend Max Born telling him that, together with Nathan Rosen, he had arrived at the interesting result that gravitational waves did not exist, though they had been assumed a certainty to the first approximation. Einstein concluded that this showed that the non-linear general relativistic field equations could tell us more or rather limit us more than we had believed up to then (Einstein and Born 1969, Einstein to Born, 1936, Letter 71, undated).

In 1936 Einstein, together with Rosen, tried to solve the non-linear field equations and find exact plane gravitational waves, but he then had to introduce singularities into the components of the metric describing the wave. However, Einstein objected to singularities, and in light of the singularities he attempted to demonstrate by an argument that no exact plane gravitational waves solutions to his field equations exist.

In June 1936, Einstein and Rosen sent the paper on gravitational waves, "Do Gravitational Waves Exist?" to *The Physical Review*. However, *The Physical Review* rejected the paper, provoking Einstein's furious reaction. The anonymous referees report on Einstein's paper included a

4discussion of the fact that (coordinate) singularities in the metric are unavoidable when describing plane gravitational waves with infinite wave fronts.

Einstein wrote to the editor of the *The Physical Review* in German that he and Rosen had sent him their manuscript for publication and had not authorized him to show it to specialists before it was printed. Einstein told the editor he therefore saw no reason to address the erroneous comments of his anonymous expert, and he consequently preferred to publish the paper elsewhere. He added that Rosen, who had in the meantime left for the Soviet Union, had authorized him to represent him in that matter. The anonymous *Physical Review* referee was the cosmologist Harvey Percy Robertson (Kennefick 1999, 210).

After this incident Einstein never published again in *The Physical Review*. He instead submitted the paper to the *Journal of the Franklin Institute* in which he had already published a paper the previous year.

In 1936 Leopold Infeld arrived in Princeton to replace Rosen as Einstein's new assistant. In his autobiography Infeld describes his first meeting with Einstein, at which Einstein explained to him his proof of the non-existence of gravity waves. Einstein began to talk about his latest and still unpublished paper concerning the work done with his assistant Rosen during the preceding year. It was on the problem of gravitational waves. Infeld explains the basic idea in simple words in the following way (Infeld 1941, 260-261): general relativity is a field theory, and it does for the problem of gravitation what Maxwell's theory had done for the problem of electromagnetic phenomena. For this reason, gravitational waves can be deduced from general relativity just as the existence of electromagnetic waves can be deduced from Maxwell's theory. In their motion the stars send out gravitational waves, spreading in time through space, just as oscillating electrons send out electromagnetic waves. It is a common feature of all field theories that the influence of one object on another spreads through space with a great but finite velocity in the form of waves.

Einstein always believed that a more thorough examination could only confirm this result, revealing some finer features of the gravitational waves. However, in the previous two years – before Infeld's arrival at Princeton in 1936 – Einstein began to doubt the existence of the gravitational waves. When making an approximate investigation of the problem he found that gravitational waves seemed to exist. But a deeper analysis brought him to the conclusion that completely contradicted the previous conclusion. Einstein thought that if this result that gravitational waves did not exist was true, it would be of a fundamental nature, because in this case, unlike previous beliefs, a filed theory (general relativity) could not then be closely connected with the existence of waves.

Einstein asked Infeld to accompany him home, where he would give him the manuscript of his paper. On the way they talked physics. Einstein spoke on the subject of gravitational waves, to which they returned many times in their conversations later. Infeld went home with the manuscript of Einstein and Rosen's paper.

Infeld was skeptical about this latest result. Although he admired Einstein as the greatest scientist in the world, he still trusted his own brain more than his admiration for Einstein; and he could not accept the nonexistence of gravitational waves. His own intuition did not let him accept this latest result dogmatically.

On the same day that Infeld had his talk with Einstein he met Robertson, of whose work on general relativity and cosmology he was well aware. Robertson was a professor of theoretical physics at Princeton who had just returned from a sabbatical leave at Caltech. Infeld told Robertson about Einstein's new gravitational waves paper that Einstein had given him to read and which he had not finished reading but felt that the result still seemed strange to him. Robertson exclaimed right away that he also did not believe in the result and said there must be a mistake somewhere in Einstein's paper. Gravitational waves do exist. He was sure about this. Infeld agreed with Robertson's judgement and they continued their discussion for a long time in Robertson's office.

Infeld carefully studied Einstein's gravitational waves paper after the meeting with Robertson and was very much impressed with this manuscript leading to the conclusion that gravitational waves do not exist! It would seem that in the long run Infeld still trusted his admiration for Einstein more than anything else.

After Infeld again had a talk with Einstein, he met Robertson the next day and told him that he had become convinced that gravitational waves do not exist. Infeld was even convinced that he was able to demonstrate this, but Robertson dismissed the idea. He took the two pages on which Infeld wrote his idea, checked all the steps of the argument, and claimed there must be a mistake in his calculations. Indeed, he found a trivial mistake: Infeld had put a minus instead of a plus. Infeld discussed gravitational waves further with Robertson and these discussions convinced him that gravitational waves do indeed exist. But if that is true, there must be after all a mistake in Einstein's paper.

In their next meeting Robertson clarified for Infeld the mistake in Einstein's explanation on gravitational waves: The linearized approximation does indeed lead to plane transverse gravitational waves. One cannot, however, exactly describe gravitational waves without introducing singularities into the components of the metric describing the wave, but these singularities are coordinate singularities and not real singularities. One can, however, deal with these singularities by a change of coordinates. Robertson, therefore, suggested effecting a "trick". He suggested that the so-called Einstein-Rosen metric (from Einstein and Rosen's paper) be transformed from space-time coordinates, suitable for representing plane gravitational waves, to cylindrical coordinates. The singularity can be located at the origin of the cylindrical axis, where one would expect to find the source of the cylindrical waves; this way the singularity can be regarded as describing a material source. The solution obtained can be considered to describe cylindrical gravitational waves rather than plane gravitational waves.

The next day Infeld went to Einstein and told him that he (Infeld) had found a mistake in the calculation, and that he believed that gravitational waves do exist. Einstein replied that he too had found a mistake in his paper with Rosen that had been submitted to *The Physical Review*. It was less trivial than Infeld's mistake in the two pages where he had tried to prove that gravitational waves do not exist, and more difficult to detect. Einstein had come to the same conclusion as Infeld's, namely that gravitational waves do in fact exist; and with Robertson's help (still not knowing it was Robertson who had reviewed and remarked on Einstein's submission to *The Physical Review*) he finally corrected his Einstein-Rosen submission paper (Weinstein 2015, 261-264).

Since Rosen had departed for the Soviet Union, Einstein acted alone in promptly and thoroughly revising their joint paper and added a section: "Rigorous Solution for Cylindrical Waves" (Einstein and Rosen 1937, 49). The new version of the paper was re-titled "On Gravitational Waves", and following Robertson's suggestion of a transformation to cylindrical coordinates, Einstein obtained exact cylindrical wave solutions of the field equations of general relativity. The metric of these waves satisfied three exact equations, the first of which, a linear equation, represented cylindrical waves in three dimensional Euclidean space (the field is independent of $x_4$). Hence Einstein presented cylindrical waves that are locally the same as plane waves (Einstein and Rosen 1937, 52-53).

Einstein concluded his paper by saying that a progressive wave can be represented with good approximation by a quantity that cannot vanish and always has the same sign. Progressive waves therefore produce a secular change in the metric. This is related to the fact that the waves transport energy, which is bound up with a systematic change in time of a gravitating mass (in effect a source of the gravitational waves) localized in the (origin) axis *x = 0*. Einstein, therefore, represented matter (the source of the gravitational waves) by singularities of the field (Einstein and Rosen 1937, 54).

This is the version which eventually appeared in the 1937 *Journal of the Franklin Institute*. The irony is that Einstein could have found the above escape to cylindrical waves months before, simply by reading *The Physical Review* referee's report, which he had dismissed so hastily (Kennefick 2005, 43-48).

At the end of the paper, "On Gravitational Waves", Einstein added a note that the second part of the paper on cylindrical waves had been considerably altered by him, after the departure of Rosen for Russia, because they had originally interpreted their formula results erroneously. He said he wished to thank his colleague, Professor Robertson, for his friendly assistance in the clarification of the original error (Einstein and Rosen 1937, 54).

In 1938 Einstein, Infeld and Banesh Hoffmann wished to create a unified field theory that would encompass both gravity and electromagnetism. The problem was that ordinary Maxwell equations for empty space were field equations that were *linear*, in which electrical particles were regarded as point singularities of the field. However, the motion of these singularities was not determined by these linear field equations. Moreover, the vacuum field equations of general relativity were *nonlinear*, and they determined the motion of the material points represented as singularities in the field.

There are three possible approximations when approaching the task of solving the Einstein field equations: the gravitational field is weak, it is static and material particles are moving slowly. In 1916 and 1918, Einstein considered the gravitational field to be weak and like the equations of electromagnetism to be linear. This approximation does not limit the acceleration of the material particles and indeed accelerating material points produce gravitational waves.

In 1938 Einstein proposed a new method of approximation for determining the gravitational field of a moving particle – choose a weak field approximation and consider very low accelerations. In the 1938 paper with Infeld and Hoffmann, Einstein therefore considered the weak field approximation and put a limit to the acceleration of the material particles. This is called, the post-Newtonian approximation.

Einstein with his assistants, Infeld and Hoffmann, calculated the first two stages of this approximation and found that in the first stage the equations of motion take the Newtonian form (Einstein, Infeld and Hoffmann 1938, 65-66). In this approximation, if we consider very low accelerations then the exact equations of motion indeed take the Newtonian form and we obtain a material particle that cannot radiate. In this state of affairs, we have revived the good old assumption that there could be no gravitational waves.